\documentclass[11pt]{article}
\usepackage[scaled=.92]{helvet}
\usepackage{graphicx}
\usepackage{booktabs}
\usepackage{listings}
\usepackage[printonlyused,nohyperlinks]{acronym}
\usepackage{color}
\usepackage{comment}
\usepackage[sort&compress]{natbib}
\usepackage{tikz}
\usetikzlibrary{positioning,shadows,arrows, fit}
\usepackage[compact]{titlesec}
\usepackage{color}
\usepackage{amsmath,amsthm,amssymb,hyperref, amsfonts}
\usepackage{dsfont}
\usepackage{threeparttable}
\usepackage{epsfig}
\usepackage{booktabs}
\usepackage{multirow}
\usepackage{longtable}
\usepackage{tabularx}
\usepackage{tocvsec2}
\usepackage{colortbl}
\usepackage{float}
\usepackage{enumerate}
\usepackage{authblk}
\usepackage[titletoc,title]{appendix}

\allowdisplaybreaks

\definecolor{greytext}{gray}{0.5}
\newcommand{\bs}{\boldsymbol}
\newcommand{\bc}{\begin{center}}
\newcommand{\ec}{\end{center}}

\newcommand{\iid}{\stackrel{\mathrm{iid}}{\sim}}
\newcommand{\mb}{\mathbf}

\def\Cov{{\rm Cov}\,}

\begin{document}
\title{Bayesian Melding of the Dead--Reckoned Path and GPS Measurements for an Accurate and
High--Resolution Path of Marine Mammals}
\author[1]{Yang Liu\thanks{Corresponding Author: yang.liu@stat.ubc.ca}}
\author[2]{Brian C. Battaile}
\author[1]{James V. Zidek}
\author[2]{Andrew W. Trites}
\affil[1]{Department of Statistics, University of British Columbia}
\affil[2]{Marine Mammal Research Unit, Fisheries Centre, AERL, University of British Columbia}
\maketitle
\begin{abstract}
With recent advances in electrical engineering, devices attached to free--ranging marine mammals today can collect oceanographic data in remarkably high spatial--temporal resolution. However, those data cannot be fully utilized without a matching high--resolution, accurate path for the animal, something that is currently missing in this field. In this paper, we develop a Bayesian Melding approach based on a Brownian Bridge process to combine the fine-resolution but seriously biased Dead--Reckoned path and the precise but sparse GPS measurements; the result is an accurate and high--resolution estimated path together with credible bands as quantified uncertainty statements. We also exploit the properties of underlying processes and some approximations to the likelihood to dramatically reduce the computational burden of handling those big high--resolution data sets.
\end{abstract}

\section{Introduction}
The idea of using free--ranging marine animals as platforms to collect oceanographic data, such as temperature and salinity, can be traced back to the discussion in~\citet{evans1970uses}, but only recent advances in electrical engineering make the animal--borne sensors feasible. Miniaturized sensors (tags) can now be attached to the animal and relay data about the environment as well as an animal's movements, behavior, physiology, something which is usually referred to as ``bio--logging"~\citep{Rutz2009}. The oceanographic data collected in bio--logging has successfully filled in the some ``blind-spots" in parts of the oceans where little or no other data are currently available~\citep{Boehme2010}. For example, animal--borne sensors can collect data at high latitudes or during the winter, where few 
ship--based measurements are available. When compared to the Argo\footnote{\url{http://www.argo.ucsd.edu/About_Argo.html}} floats, which can also work in those situations, the animal-borne sensors can take measurements in the ocean currents or upwelling zones, where the Argo floats will drift away. We also need have little concern that animal-borne sensors can crash into sea floor or ice, since the animals, unlike Argo floats, can ``automatically" avoid those obstacles.

There are numerous examples showing how data collected in bio--logging can help improve our knowledge of the ocean's environment. \citet{Lydersen2002} first deployed sensors on white whales to monitor the salinity and temperature structure of an ice--covered Arctic fjord. The data collected by Southern elephant seals helped to identify the Antarctic circumpolar current fronts in the South Atlantic~\citep{Boehme2008} and the seasonal evolution of the upper--ocean adjacent to the South Orkney Islands, Southern Ocean~\citep{Meredith2011}.  \citet{Isachsen2014} combined Argos floats data with the data collected by hooded seals and so revealed that the Nordic Seas are getting warmer and saltier. Besides the ocean hydrographic data, the originally poorly sampled bathymetry map of the Antarctic continental shelf was improved using data collected by elephant seals~\citep{Padman2010}. More examples on the contributions of animal--borne sensors can be found in~\citet{Boehme2010}.

In most of the studies above, the data were first transmitted to a satellite by the sensor and then downloaded from the satellite for analysis. Due to limitations in the communication bandwidth and cost, a very small amount of data can be transferred in this work--flow and usually we can only obtain data with a low spatial--temporal resolution. For example in~\citet{Boehme2008}, only two temperature and salinity profiles\footnote{Profile here means a sequence of measurements at different depths} are obtained per day and the spatial resolution of these data are 20--50km. Such low resolution data markedly restricts the scope and contribution of the scientific findings from the studies with animal--borne sensors~\citep{Boehme2008}.

Fortunately some species of animals carrying the sensors, like elephant seals or fur seals, have a relatively fixed ``home" (breeding habitat) on an island or sea shore, which is usually the start and end points of their foraging trips. Most sensors deployed in bio--logging are actually attached to the animals on those islands~\citep[see e.g.][]{Boehme2008, Nordstrom2013a}, where they can be also retrieved when the animals return. It is thus possible to circle around the bottle neck of satellite communication by storing the data locally on the sensor and then download data after they are retrieved. This approach can provide a nearly continuous record of the sea water temperature as well as variables related the animal's movement and behavior, e.g.~diving depth, acceleration etc., with temporal resolution of 1 second or even higher~\citep[see e.g.][]{Wilson2008, Nordstrom2013a}. 

Such high--resolution bio--logging data has contributed much to our knowledge of animal's behavior and movement pattern~\citep{Dowd2011} and energy expenditure~\citep{Wilson2007}, but few studies have taken advantage of this high--resolution record of water temperature to achieve more knowledge about the ocean environment or the relationship between the tracked animal and the environment~\citep{Nordstrom2013a}. The main reason here is the lack of an accurate and high--resolution path (in longitude and latitude) of the tagged animal so we cannot know where the temperature is measured. The same reason also limits our understanding of the relationship between the animal's habitat preference and the environment~\citep{Benoit-Bird2013,Benoit-Bird2013a}. 

Currently in bio--logging, an animal's location is usually determined by the direct measurements from Global Positioning System~\citep[GPS][]{hofmann1993global}, which can only work when the sensor floats on the water surface and has direct lines of sight to four or more satellites. However, the marine mammals usually spend very little time on the sea surface and the GPS sensor only provides sparse and irregular observations of the animal's path. Some studies today impute the gaps between two GPS observations via linear interpolation~\citep{Benoit-Bird2013,Nordstrom2013a} or a statistical model fitted to the GPS observations. Examples of the latter approach include the correlated random walk~\citep{Jonsen2005} and continuous time correlated random walk~\citep{Johnson2008}. Both the linear interpolation and the more sophisticated models impose certain parametric assumptions that may incur bias in the imputed path.

In order to fill in the gaps between GPS observations, a ``Dead--Reckoning" (DR) tag, which usually consists of an accelerometer, a magnetometer, a time--depth--recorder (TDR) and other supporting components, is also deployed on the instrumented animals~\citep[see e.g.][]{Wilson2007,Nordstrom2013a}. This DR tag can sample at infra--second frequencies, like 16Hz or 32Hz, so a detailed record of the animal's movement can thus be obtained. After retrieving the tag, we can download the data and use the so--called ``Dead--Reckoning Algorithm"~\citep[DRA,][]{Wilson2007,Elkaim2006} to reconstruct an estimated path from the DR tag data. The basic idea of DRA is as follows: estimate the animal's orientation (direction of velocity) from the accelerometer and magnetometer readings by solving the Wahba's problem~\citep{wahba1965least}; next estimate the animal's speed (norm of the velocity) using the TDR data or by assuming a known constant; estimate the path by integrating the velocity from a known starting point. This estimated path is called the ``Dead--Reckoned'' (DR) path hereafter.

The temporal resolution of the DR path is decided by the sampling frequency of the DR tag, namely, 1/16 or 1/32 of a second and the spatial resolution of it can be as fine as a few meters. The DR path provides remarkably detailed information about the animal's movement, especially fine scale fluctuations that the GPS is not able to capture. However, the DR path can be seriously biased due to measurement error or systematic bias in the DR tag, assumptions and errors in the orientation and speed estimation, and discretization bias in the integration, etc~\citep{Wilson2007}. A more detailed review of the DRA is provided in Appendix~\ref{ap:DRA}. As shown later in our case study, the bias of the DR path can be as large as 100km at the end of a seven--day trip. Therefore, we must correct the bias in the DR path before applying it, to provide an in--situ record of the oceanographic variables as well as to address other biological or ecological questions. 

Fortunately, extensive tests have demonstrated the unbiasedness and high precision of the GPS observations with standard errors around a couple of hundred meters~\citep{Bryant2007,Hazel2009} so they can be used to correct the DR path. The conventional correction method in~\citet{Wilson2007} can be summarized as follows: denote the DR path (in one dimension) as $x_1, x_2, \ldots, x_T$ at times $t=1, 2, \ldots, T$ and the GPS observations at times $1$ and $T$ are $y_1, y_T$ respectively; assume without loss of generality, that $x_1 = y_1=0$ and that the corrected path $\hat{\eta}_t$ is calculated as, 
\begin{align}
\hat{\eta}_t = x_t + \frac{y_T - x_T}{T-1}(t-1), \label{eq:adhoc-DRABiasCorrection}
\end{align}
which evenly distributes the bias $y_T - x_T$ evenly over the individual time points. The DR path between two GPS observations is directly shifted to the locations indicated by the GPS observations, namely $\hat{\eta}_1= y_1$ and $\hat{\eta}_T= y_T$. This procedure is repeated for all the sections separated by the GPS observations to correct the whole path. \citet{Wilson2007} did not provide any justification for this conventional correction method. Also, it fails to take account of the measurement error in the GPS observations, nor does it provide an uncertainty statement about the corrected path. According to~\citet{Nordstrom2013a}, the bio--logging community has concern about the validity of the corrected path and few applications are developed based on it. 

In this paper, we use the Bayesian Melding (BM) approach to develop a statistically rigorous method for correcting the bias in each of the geographical coordinates of the DR path. That approach was pioneered by~\citet{Fuentes2005} to combine the direct observations of air--pollutant level from a sparse network of monitoring stations and the output from deterministic chemical transportation (computer) model outputs, at each pixel of a map based on known pollutant source and geophysical information. The BM approach was later adapted for use in a variety of different fields, such as modeling hurricane surface wind~\citep{Foley2008}, ozone level~\citep{Liu2011}, and wet deposition~\citep{Sahu2010}, etc. All these applications have demonstrated the remarkable flexibility and effectiveness of the BM approach. 
When comparing our application to that in~\citet{Fuentes2005}, the GPS observations play the role of the station measurements while the DR path plays the role of the computer model output. Using the GPS to correct DR path can also be viewed as combining the location information from the GPS and DR path, which is the very strength of BM.  For each coordinate the method provides Bayesian credible intervals to quantify the residual uncertainty about the corrected coordinates. The estimated path can then be found by combining in the obvious way, the separate curves for the two geographical coordinates. 

Our BM model assumes the animal's path is a Brownian Bridge process and the GPS observations are unbiased observations of this true process with $i.i.d.$ normal measurement errors. The DR path is assumed to be the sum of the true process, a systematic error component that is modeled by a parametric function, and a random error component that is modeled by a Brownian motion process. As discussed in the sequel, these model choices are supported by the biological and ecological literature. 

All components in our BM model are Gaussian and linear, which makes inference conceptually easy. However in seeking to retain the benefits of the high--resolution DR path we encounter a big data problem. For example, a typical one--week foraging trip in our case study, with 16Hz sampling frequency, resulted in $T = 7 \times 24 \times 3600 \times 16 = 9,676, 800$ time points. The large $T$ rules out the decomposition and storage of the covariance matrix of the $T$--dimensional true process on most computers. But we show how that high computational burden 
we can be dramatically reduced by coupling the conditional independence properties of the Brownian Bridge and Brownian Motion processes with some judicious approximations to the likelihood. The computations needed to implement the resulting BM model can be completed on a regular laptop in under five minutes.

To illustration the capability of our BM approach, we apply it to data from two foraging trips of northern fur seals in the summer of 2009 in the Bering sea, 
observed as part of the Bering Sea Integrated Research Program (BSIERP)~\citep{Nordstrom2013a,Benoit-Bird2013}. Cross--validation studies carried out for model selection show our method to be superior to the conventional bias correction method in \citet{Wilson2007}. 

The paper is organized as follows. Section 2 introduces our BM model. Section 3 describes how model inference can be carried out efficiently. The real data application together with cross--validation studies is found in Section 4. Section 5 concludes this paper and discusses some future work.

\section{Bayesian Melding model} \label{sc:BM}
In this section, we adopt the BM approach to combine the information from the accurate but sparse GPS observations and biased but dense DRA results. For simplicity, the two dimensions of the path (latitude and longitude) are dealt with separately. Thus abstractly we are considering a one dimensional path over time 
which we denote by $\eta(t)$ at discrete time points $t=1, 2, \ldots, T$. The time unit plays no essential role in the mathematical development our theory. Moreover the approach works just as well with unequally spaced time points, in other words, for arbitrary $t_1, t_2, \ldots, t_T$. But for expository simplicity we work with $1:T$ as the real bio--logging data are equally spaced. 
As in the previous BM literature, we take $\eta(t)$ to be a Gaussian process, with\begin{align}
\bs{\eta}(1:T) \sim N(\mb{f}(1:T), \mb{R}(1:T, 1:T)), \label{eq:BM-etaModel}
\end{align}
where the $\mb{f}(\cdot)$ denotes the process mean function and $\mb{R}$, its covariance matrix. The notation $\mb{f}(1:T)$ stands for the vector $\{f(1), f(2), \ldots, f(T) \}^T$ while $\mb{R}(1:T, 1:T)$ is a $T \times T$ covariance matrix with $R(t, t') = \Cov(\eta(t), \eta(t'))$. Throughout this paper, bold faced characters are used exclusively to represent vectors or matrices. 

\begin{figure}[h]
\includegraphics[width=\textwidth]{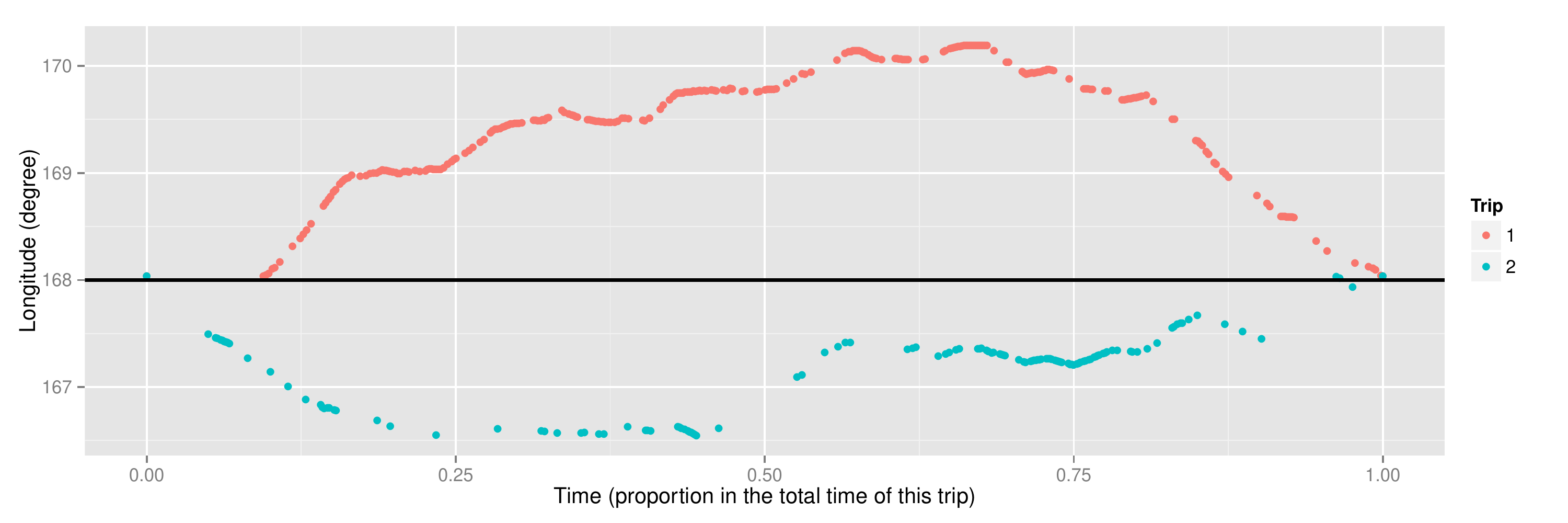}
\caption{The longitude in the GPS observations of the two trips in our case study of northern fur seals. Both trips started and ended at Bogoslof Island and the horizontal line indicates the longitude $-168.035^\circ E$ of the island. The time unit in this figure is the proportion of the total time of this trip.} \label{fg:egBrownianBridge}
\end{figure}

Various choices are available for this Gaussian process. A common one~\citep{Sacks1989, Fuentes2005} assumes that $\mb{f}$ to be a simple parametric model, e.g., a constant or a linear function of the covariates, and $R(t, t') = \sigma^2 \rho(|t-t'|)$, where $\rho(\cdot)$ is an isotropic correlation function from a class such as the ~Mat\'ern or power exponential. However, this popular stationary Gaussian process is not suitable for our application. As noted above, high--resolution data can only be obtained by retrieving the tag, which means that the start and end points of the path are fixed (usually the same location), which is illustrated in Figure~\ref{fg:egBrownianBridge}. Apart for the start and end points, the path is unknown, and hence random in our Bayesian framework. Its variation is relatively large in the middle and small when close to the known start and end points. These features of the path inspire us to model it with a Brownian Bridge process, whose mean and covariance functions are:
\begin{align*}
f(t) =& A + (B-A)\frac{t-1}{T-1}\\
R(s, t) =&\sigma^2_H \frac{(\min(s, t)-1)(T-\max(s, t))}{(T-1)}
\end{align*}
where $\eta(1)=A$ and $\eta(T)=B$ are the known start and end points of the path, while $\sigma^2_H$ is the variance parameter. Notice that $R(1, \cdot) = R(\cdot, T)=0$, in accordance with the known start and end points $\eta(1)$ and $\eta(T)$. Also, $R(t, t)$ increases with $t$ when $t < (T-1)/2$ and decreases with $t$ for  $t > (T-1)/2$, reflecting the fact that the variation of the path is large in the middle. Another noteworthy property of our covariance matrix $\mb{R}$ is its form as the product of a scalar $\sigma^2$ and a matrix, the latter depending only on the time points. To clearly represent the parameters of the Brownian Bridge process, we introduce the notation 
\begin{align}
\mathrm{BB}(A, B, T_S, T_E, \sigma^2) \label{eq:BM-BBdefinition}
\end{align}
for a Brownian Bridge process, which starts from $A$ at time $T_S$ and ends in $B$ at time $T_E$ with a variance parameter $\sigma^2$, namely $f(t) = A + (B-A)\frac{t-T_S}{T_E-T_S}$ and covariance function $R(s, t) =\sigma^2_H \frac{(\min(s, t)-T_S)(T_E-\max(s, t))}{(T_E-T_S)}$.

Our choice of the Brownian Bridge model is well supported in the biology and ecology literature. According to~\citet{Humphries2010}, marine mammals tend to exhibit Brownian--like movements in environments with abundant food resources, such as the ocean around Bogoslof island where our case study is centered, one believed to be just such a resourceful environment~\citep{Benoit-Bird2013a}. Also, a Brownian Bridge model was proposed~\citet{Horne2007} to model the habitat use for a wide range of animal species. This model is well accepted by the biology and ecology field and further improved by~\citet{Sawyer2009},~\cite{Kranstauber2012}, and~\citet{Kranstauber2014}. Many other examples of modeling an animal's path with Brownian Bridge processes can be found in the references of the above papers.

The GPS observations of the locations are denoted by $Y(t_k), k=1, 2, \ldots, K, t_1=1, t_K=T, t_k \in \{2, \ldots, T-1\}, k=2, \ldots, K-2$, which are unbiased observations of the true location:
\begin{align}
Y(t_k)|\eta(t_k) \iid N(\eta(t_k), \sigma^2_G), \label{eq:BM-GPSModel}
\end{align}
for $ k=2, \ldots, K-2$. The known start and end points assumption implies that $Y(t_1) = \eta(t_1)$, and $Y(t_K) = \eta(t_K)$ are known.

Next, we use $X(t), t=1, 2, \ldots, T$ to denote the DR path without any error correction. To incorporate the bias of the DR path, we assume:
\begin{align}
X(t) = \eta(t) + h(t)+ \xi(t), \label{eq:BM-DRModel}
\end{align}
where $h(t)$ is a parametric bias term designed to capture 
trend if any in the DRA path
while $\xi(t)$ denotes another Gaussian process independent of $\eta(t)$
that captures any irregular components in the deviation of 
the DR path from the truth:
\begin{align*}
\bs{\xi}(1:T) \sim N(\mb{0}, \mb{C}(1:T, 1:T)).
\end{align*} 
For the parametric bias component $h(t)$, we have considered various models, e.g.~$h(t)= \sum_{i=1}^Q \beta_i t^{i-1}$. The residual bias $\xi(t)$ is assumed to be a Brownian motion process (random walk of order 1) whose covariance function is therefore
\begin{align*}
C(s, t) =&\sigma^2_D (\min(s, t)-1).
\end{align*}
We believe the Brownian motion process to be a reasonable approximation to the gradually accumulated error in the DRA. If we assume the error in the velocity estimates from the DRA at each time point is $i.i.d.$ normal, the error in the integrated path is then a Brownian motion.

The final ingredients in our BM model are the prior distributions of the parameters. For notational simplicity, all densities are denoted by square brackets $[\ldots]$ throughout this report. For $\sigma^2_G$, we assume a known constant based on the previous extensive tests of the Fastloc\textregistered~GPS device. The priors of the other two variance parameters are chosen to be the reference priors, $[\sigma^2_H] \propto \frac{1}{\sigma^2_H}$ and $[\sigma^2_D] \propto \frac{1}{\sigma^2_D}$, which is a non--informative prior on the log scale ($[\log(\sigma^2_H)] \propto 1$). For $\bs{\beta} = \{\beta_1, \beta_2, \ldots, \beta_Q\}^T$, a non--informative flat prior $[\bs{\beta}] \propto 1$ is used. All these parameters are assumed to be independent of each other.

For expository simplicity in describing the joint distribution of all the data and parameters, the following notations are introduced:
\begin{itemize}
\item The unknown part of the true path is denoted by $\bs{\eta} = \bs{\eta}(2:(T-1))$, a $T-2$ dimensional vector.
\item GPS observations of the unknown part of the path are denoted by $\bs{Y}= \{Y(t_2), Y(t_3), \ldots, Y(t_{k-1})\}^T$, a $K-2$ dimensional vector.
\item The DR path $\mb{X}= \{\mb{X}(2:(T-1)), X(T)-Y(T)\}^T$, a vector of dimension $T-1$;
\item For the two unknown variance parameters $\bs{\phi} = \{\sigma^2_H, \sigma^2_D\}^T$.
\end{itemize}

The joint likelihood of our model is 
\begin{align}
[\mb{X}, \mb{Y}, \bs{\eta}, \bs{\beta}, \bs{\phi}] = [\bs{\phi}][\bs{\beta}][\bs{\eta}|\bs{\phi}][\mb{Y}|
\bs{\eta}][\mb{X}|\bs{\beta}, \bs{\phi}, \bs{\eta}]. \label{eq:BM-JointDensity}
\end{align}
To obtain an estimate of the animal's true path and its uncertainty, we seek the posterior distribution
\begin{align}
[\bs{\beta}, \bs{\eta}|\mb{X}, \mb{Y}] = \int \underbrace{[\bs{\beta}, \bs{\eta}|\mb{X}, \mb{Y}, \bs{\phi}]}_{\textrm{(1)}} \times  \underbrace{[\bs{\phi}|\mb{X}, \mb{Y}]}_{\textrm{(2)}} \mb{d} \bs{\phi}. \label{eq:BM-TargetPost}
\end{align}
Here we also include the $\bs{\beta}$ term, which can be used to assess the bias of the DRA. The posterior mean, denoted by $\tilde{\eta}(t)$, can be an estimate of the animal's path and the posterior standard error, denoted by $\tilde{\sigma}(t)$ provides an uncertainty statement about the estimated path. The point--wise 95\% credible interval for $\eta(t)$ can be constructed as 
\begin{align}
[\tilde{\eta}(t) - 1.96\tilde{\sigma}(t), \tilde{\eta}(t) + 1.96\tilde{\sigma}(t)]. \label{eq:BM-PosteriorCI}
\end{align}
 
\section{Model Inference} \label{sc:BM-Inference}
To calculate the posterior~\eqref{eq:BM-TargetPost}, we first fix the variance parameters $\bs{\phi}$ and calculate part (1) in Equation~\eqref{eq:BM-TargetPost} and then integrate over the posterior of $\bs{\phi}$. The first part of this section introduces how the components of~\eqref{eq:BM-TargetPost} can be efficiently evaluated and the second part describes how we approximate the integral. 
A comparison of our BM approach to the conventional approach is included in Section~\ref{ss:theoryComparisonBMtoAdhoc}.

\subsection{Evaluating components of the posterior}
For notational simplicity, we use $\langle \cdot|\cdot \rangle$ to denote $[\cdot|\cdot, \bs{\phi}]$, namely $\langle \bs{\eta}|\mb{X}, \mb{Y} \rangle= [\bs{\eta}|\mb{X}, \mb{Y}, \bs{\phi}]$. As we specify our model in a Gaussian and linear fashion, it is straightforward to show that $\langle \bs{\beta}, \bs{\eta}|\mb{X}, \mb{Y} \rangle$ is a multivariate Gaussian density,
\begin{align}
\langle \bs{\beta}, \bs{\eta}|\mb{X}, \mb{Y}\rangle  \propto \exp\left\{ -\frac{1}{2} \left[ (\bs{\zeta}-\mb{M}^{-1}_1 \mb{M}_2)^T \mb{M}_1 (\bs{\zeta}-\mb{M}^{-1}_1 \mb{M}_2) \right] \right\}
 \label{eq:BM-Zeta-cPhi}
\end{align}
where $\bs{\zeta}=\{\bs{\beta}^T, \bs{\eta}^T\}^T$ and $\mb{M}_1, \mb{M}_2$ are derived in Appendix~\ref{ap:betaEta_Phi}. Although the multivariate Gaussian posterior makes the inference conceptually easy, calculating its posterior mean $\mb{M}_1^{-1} \mb{M}_2$ and covariance matrix $\mb{M}_1^{-1}$ actually involves a matrix decomposition with computational complexity of order $O(T^3)$, which is a tremendous computational burden when $T$ is large. It is possible to avoid the $O(T^3)$ matrix decomposition with certain sparse matrix techniques together with the Sherman--Morrison--Woodbury formula~\citep{henderson1981deriving}, but those techniques still require the storage of some huge matrices and complicated matrix calculations. This pushes us to further reduce the complexity of~\eqref{eq:BM-Zeta-cPhi}.

It is easily seen that we have more information (data) about 
$\bs{\eta}_G \triangleq \bs{\eta}(t_{1:K})$ 
where the GPS observations are available than where they are not. For 
$\bs{\eta}(1:T\setminus t_{1:K})$, we only have the DR path. So our first step is to break 
$\bs{\eta}$ into two sets, that is
\begin{align}
\langle \bs{\beta}, \bs{\eta}|\mb{X}, \mb{Y} \rangle = \langle \bs{\eta}(1:T\setminus t_{1:K})|\bs{\beta}, \bs{\eta}_G, \mb{X}, \mb{Y} \rangle \langle \bs{\beta}, \bs{\eta}_G, \mb{X}, \mb{Y} \rangle. \label{eq:BM-cPhi-b1}
\end{align}
We can then use the Markovian property of the Brownian Bridge process~\citep[see e.g.][]{stirzaker2001probability} to simplify~\eqref{eq:BM-cPhi-b1} as:
\begin{align}
\langle \bs{\beta}, \bs{\eta}|\mb{X}, \mb{Y}, \bs{\phi} \rangle = \left\{ \prod_{k=1}^{K-1} \langle\bs{\eta}(t_k +1 : t_{k+1} - 1)|\eta(t_k), \eta(t_{k+1}), \bs{\beta}, \mb{X}, \mb{Y} \rangle \right\} \times \langle \bs{\beta}, \bs{\eta}_G|\mb{X}, \mb{Y} \rangle. \label{eq:BM-cPhi-b2}
\end{align}
In this way, we partition the long $\bs{\eta}$ series into small pieces separated by the GPS observations. We also exploit the Markovian property of the Brownian Motion and find the $k^{\rm th}$ term in the first part of~\eqref{eq:BM-cPhi-b2} can be simplified as
\begin{align}
\langle \bs{\eta}(t_k +1 : t_{k+1}-1)|& \eta(t_k), \eta(t_{k+1}), \bs{\beta}, \mb{X}, \mb{Y} \rangle = \notag \\
&\langle \bs{\eta}(t_k +1 : t_{k+1}-1)| \eta(t_k), \eta(t_{k+1}), \bs{\beta}, \mb{X}(t_k : t_{k+1})\rangle. \label{eq:BM-cPhi-b3}
\end{align}
All the derivations for~\eqref{eq:BM-cPhi-b2} and~\eqref{eq:BM-cPhi-b3} are provided in Appendix~\ref{ap:condIndep}. In~\eqref{eq:BM-cPhi-b3}, the posterior of $\eta(t)$ between two GPS points can be evaluated only with the corresponding DR path together with the posterior distribution of the two GPS points and $\bs{\beta}$. This remarkably simple change greatly reduces the memory cost when computing the posterior of the long sequence and enables us to easily parallelize the whole calculation. Moreover, both Brownian Bridge and Brownian Motion processes conditioned on two end points are Brownian Bridge processes, such that,
\begin{align*}
\bs{\eta}(t_k +1 : t_{k+1}-1)| \eta(t_k), \eta(t_{k+1})\sim & \mathrm{BB}(\eta(t_k), \eta(t_{k+1}), t_k, t_{k+1}, \sigma^2_H) \\
\bs{\xi}(t_k +1 : t_{k+1}-1)| \xi(t_k), \xi(t_{k+1}) \sim & \mathrm{BB}(\xi(t_k), \xi(t_{k+1}), t_k, t_{k+1}, 
\sigma^2_D).
\end{align*} 
This fact is exploited to completely avoid the matrix inverse calculation when evaluating~\eqref{eq:BM-cPhi-b3}, which further reduces the computational burden. The derivations are included in Appendix~\ref{ap:BM-perCond}.  

However, when evaluating $ \langle \bs{\beta}, \bs{\eta}_G|\mb{X}, \mb{Y}\rangle$ in~\eqref{eq:BM-cPhi-b2}, we still need to deal with the long sequence $\mb{X}$. However, the $\mb{Y}$ is an unbiased observation of $\bs{\eta}_G$ and therefore 
$\langle \bs{\eta}_G|\mb{X}, \mb{Y}\rangle$ can be well approximated by $\langle \bs{\eta}_G|\mb{Y} \rangle$. This approximation is exceptionally good when $\sigma^2_D > \sigma^2_G$. For $\bs{\beta}$, it can be well inferred from the difference between $\mb{X}_G \triangleq \mb{X}(t_{1:K})$ and $\mb{Y}$. In this way, we introduce the following approximation:
\begin{align}
\langle \bs{\beta}, \bs{\eta}_G|\mb{X}, \mb{Y}\rangle \approx \langle \bs{\beta}, \bs{\eta}_G|\mb{X}_G, \mb{Y}\rangle. \label{eq:BM-cPhi-ApproxGBeta}
\end{align}
With similar arguments, we also approximate the posterior of $\bs{\phi}$ by,
\begin{align}
[\bs{\phi}|\mb{X}, \mb{Y}]  \approx [\bs{\phi}|\mb{X}_G, \mb{Y}] \label{eq:BM-ApproxPhi} 
\end{align}
The explicit expressions for~\eqref{eq:BM-cPhi-ApproxGBeta} and~\eqref{eq:BM-ApproxPhi} are included in Appendix~\ref{ap:betaEtaGPhi_XGY}. Our simulations which are designed to mimic the real data sets have shown that the impact of the two approximation errors in~\eqref{eq:BM-cPhi-ApproxGBeta} and~\eqref{eq:BM-ApproxPhi} is negligible.

In summary, the posterior of $\bs{\eta}$ is approximated as follows:
\begin{align}
[\bs{\eta}, \bs{\beta}|&\mb{X}, \mb{Y}] = \int[\bs{\eta}, \bs{\beta}|\mb{X}, \mb{Y}, \bs{\phi}] [\bs{\phi}|\mb{X}, \mb{Y}] d \bs{\phi} \notag \\
= \int& \langle \bs{\eta}(1:T\setminus t_{1:K})|\bs{\beta}, \bs{\eta}_G, \mb{X}, \mb{Y} \rangle  
\langle\bs{\eta}_G, \bs{\beta}|\mb{X}, \mb{Y} \rangle [\bs{\phi}|\mb{X}, \mb{Y}] d \bs{\phi} \notag \\
= \int&\left\{ \prod_{k=1}^{K-1} \langle \bs{\eta}(t_k +1 : t_{k+1} - 1)|\bs{\beta}, \eta(t_k), \eta(t_{k+1}), \mb{X}\rangle \right\} 
\langle \bs{\eta}_G, \bs{\beta}|\mb{X}, \mb{Y} \rangle [\bs{\phi}|\mb{X}, \mb{Y}] d \bs{\phi} \notag \\
= \int&\left\{ \prod_{k=1}^{K-1} \langle \bs{\eta}(t_k +1 : t_{k+1} - 1)|\bs{\beta}, \eta(t_k), \eta(t_{k+1}), \mb{X}(t_k: t_{k+1})\rangle \right\} 
\langle \bs{\eta}_G, \bs{\beta}|\mb{X}, \mb{Y} \rangle [\bs{\phi}|\mb{X}, \mb{Y}] d \bs{\phi} \notag \\
\approx \int&\left\{ \prod_{k=1}^{K-1} \langle \bs{\eta}(t_k +1 : t_{k+1} - 1)|\bs{\beta}, \eta(t_k), \eta(t_{k+1}), \mb{X}(t_k: t_{k+1})\rangle \right\} \langle \bs{\eta}_G, \bs{\beta}|\mb{X}_G, \mb{Y} \rangle [\bs{\phi}|\mb{X}_G, \mb{Y}] d \bs{\phi}. 
\label{eq:BM-JointPostExpression} 
\end{align}
Next, we will carry out the integration in Equation~\eqref{eq:BM-JointPostExpression}. 

\subsection{Integration over the variance parameters $\bs{\phi}$}\label{ss:IntegrationOfPhi}
According to the BM literature~\citep{Liu2011}, the integration in~\eqref{eq:BM-JointPostExpression} can be carried out by MCMC. However, we need to avoid the heavy computational burden of the MCMC techniques in our application and it may not be practical to store the MCMC samples of the high dimensional parameter $\bs{\eta}$. The first alternative is to avoid the integration via the empirical Bayesian approach as in~\citet{casella1985introduction}:
\begin{align*}
[\bs{\beta}, \bs{\eta}|\mb{X}, \mb{Y}] \approx  [\bs{\beta}, \bs{\eta}|\mb{X}, \mb{Y}, \bs{\hat{\phi}}],
\end{align*} 
where $\bs{\hat{\phi}}$ is the maximum likelihood estimate of $\bs{\phi}$ after $\bs{\eta}, \bs{\beta}$ are marginalized out,
\begin{align}
\bs{\hat{\phi}} = \arg\max_{\bs{\phi}} \log([\bs{\phi}|\mb{X}_G, \mb{Y}]). \label{eq:BM-marginalMLE-Phi}
\end{align}

The empirical Bayesian approach is computationally simple, especially when we can explicitly evaluate the marginal likelihood. However, it fails to reflect the uncertainty in $\bs{\phi}$ and thus it underestimates the uncertainty in the posterior of $\bs{\eta}$. To overcome this issue, we use a numerical integration method like that in INLA (integrated nested Laplace approximation,~\citet{Rue2009}), which approximates the integration on a grid decided by the likelihood surface. 

Let $\mb{H}$ denote the $2 \times 2$ Hessian matrix of $\bs{\hat{\phi}} = \{\hat{\sigma}^2_H, \hat{\sigma}^2_D\}^T$ in~\eqref{eq:BM-marginalMLE-Phi} and $\bs{\Sigma} = \mb{H}^{-1}$.  With the eigenvalue decomposition $\bs{\Sigma} = \mb{A} \bs{\Lambda} \mb{A}^T$, the space of $\bs{\phi}$ can be explored by $\bs{\phi}(\mb{z}) = \bs{\hat{\phi}} +  \mb{A} \bs{\Lambda}^{1/2} \mb{z}$, where $\mb{z}$ is a $2 \times 1$ vector. To find the grid for numerical integration, we start from $\mb{z}=\mb{0}$ and search in the positive direction of $z_1$, that is, we increase $j \in \mathbb{N}^{+}$ and $\mb{z} = (j\delta_z, 0)$ as long as
\begin{align*}
\log([\bs{\phi}(\mb{0})|\mb{X}_G, \mb{Y}]) - \log([\bs{\phi}(\mb{z})|\mb{X}_G, \mb{Y}]) < \delta_\pi,
\end{align*} 
where $\delta_z$ is the step size and $\delta_\pi$ controls the magnitude of probability mass that will be included in the numerical integration. After searching on the positive side, we switch direction and search on the negative side of $z_1$. This procedure is repeated for both dimensions of $\mb{z}$. For our BM model above, if the search stops at $J^{+}_1, J^{+}_2$ steps in the positive directions of $z_1$ and $z_2$ respectively and $J^{-}_1, J^{-}_2$ in their negative directions, a grid of size $(J^{+}_1 + J^{-}_1 + 1) \times (J^{+}_2 + J^{-}_2 + 1)$ is used in the numerical integration and the points on this grid are $\mb{z}_{j_1, j_2}= \delta_z(j_1, j_2)$, with $j_1 \in (-J^-_1, -J^-_1+1, \ldots, 0, \ldots, J^+_1)$ and $j_2 \in (-J^-_2, -J^-_2+1, \ldots, 0, \ldots, J^+_2)$. The integral in~\eqref{eq:BM-JointPostExpression} is approximated by 
\begin{align}
[\bs{\eta}, \bs{\beta}|\mb{X}, \mb{Y}] \approx \sum_{j_1=-J^-_1}^{J^{+}_1} \sum_{j_2=-J^-_2}^{J^{+}_2} w_{j_1, j_2} \times [\bs{\eta}, \bs{\beta}|\bs{\phi}(\mb{z}_{j_1, j_2} ), \mb{X}, \mb{Y}], \label{eq:BM-etaPost-GridApprxoimation}
\end{align}
where 
\begin{align*}
w_{j_1, j_2} = \frac{[\bs{\phi}(\mb{z}(j_1, j_2))|\mb{X}_G, \mb{Y}] }{\sum_{j_1} \sum_{j_2}  [\bs{\phi}(\mb{z}(j_1, j_2))|\mb{X}_G, \mb{Y}] }.
\end{align*}
Here~\eqref{eq:BM-etaPost-GridApprxoimation} resembles Equation (5) of~\citet{Rue2009}. As the $\bs{\eta}, \bs{\beta}$ conditioning on $\bs{\phi}$ follows a multivariate Gaussian distribution, the posterior $[\bs{\eta}, \bs{\beta}|\mb{X}, \mb{Y}]$ can be approximated by a mixture of multivariate Gaussian densities, whose mean and variance can be easily calculated. The detailed expressions are in Appendix~\ref{ap:IntegrationAsNormalMixture}.

\subsection{Comparison to the conventional bias correction method} \label{ss:theoryComparisonBMtoAdhoc}
When compared to the conventional bias correction method in~\eqref{eq:adhoc-DRABiasCorrection}, our BM approach can account for both data and model uncertainty and provide a CI for the estimates of the animal's path. Moreover, there is an interesting connection between the BM posterior mean $\tilde{\eta}$ and the  conventional corrected path $\hat{\eta}$ in Equation~\eqref{eq:adhoc-DRABiasCorrection}. Without loss of generality, let $K=2$ (no GPS observations except the known start and end points) and $Y(1)=X(1) =0$. So~\eqref{eq:adhoc-DRABiasCorrection} can be written as 
\begin{align*}
\hat{\eta}(t) =Y(T) \frac{t-1}{T-1} +  \left[X(t) - X(T) \frac{t-1}{T-1} \right]. 
\end{align*}
For BM, if $h(t) =0$ for all $t$ and $\bs{\phi}$ is known, the posterior mean $\bs{\tilde{\eta}}$ under the above assumptions can be simplified into
\begin{align*}
\tilde{\eta}(t) = Y(T) \frac{t-1}{T-1} + \frac{\sigma^2_H}{\sigma^2_H + \sigma^2_D} \left[X(t) - X(T) \frac{t-1}{T-1} \right].
\end{align*}
The first parts of $\hat{\eta}(t)$ and $\tilde{\eta}(t)$ represent linear interpolations between $Y(1) =0$ and $Y(T)$, which determines the basic trend of the animal's path between two known points. The second part is a ``bridge" constructed by the $X(t)$, which starts at $X(1)=0$ and ends at $0 = X(T) - X(T) \frac{T-1}{T-1}$. This bridge can be treated as the ``detail" for the animal's path, which is then added to the basic trend of the first part. 

The difference between the conventional method and the simplified BM approach is the weight on the ``detail". In the conventional approach, the ``detail" is directly added to the basic trend while BM shrinks the detail by a factor of $\rho= \frac{\sigma^2_H}{\sigma^2_H + \sigma^2_D}$. According to our model, we cannot distinguish between $\eta(t)$ and $\xi(t)$ in $X(t)$ at those non--GPS points, as we only observe the sum of them, but we know that $\eta(t)$ accounts for the $\sigma^2_H$ part of the total $\sigma^2_H + \sigma^2_D$ variance (they are both of mean zero after $Y(T) \frac{t-1}{T-1}$ is removed). In this way, a fraction $\rho= \frac{\sigma^2_H}{\sigma^2_H + \sigma^2_D}$ of the detail is treated as signal in $\eta(t)$ and added to the basic linear trend.

Notice here we only compare the most simplified BM approach to the conventional approach. In practice, the BM $\tilde{\eta}$ is far more complicated than the form shown above with the parametric part from $\bs{\beta}$ and the integration over $\bs{\phi}$. According to our simulations and cross--validation of real data in Section~\ref{ss:BM-RealDataAna-XV}, our $\tilde{\eta}$ is a substantial improvement on the conventional estimate $\hat{\eta}$.

\section{Case studies} \label{sc:BM-RealDataAna}
The proposed BM approach was applied to two data sets collected in the 2009 Bogoslof Island northern fur seal study~\citep{Benoit-Bird2013}. These two data sets, denoted by ``Trip 1" and ``Trip 2" came from two trips of different female seals. The first trip lasted about 6 days and second trip lasted about 7 days. The GPS device in this study was programmed to make one observation attempt every 15 minutes but a large fraction of these attempts failed to obtain a valid observation of the coordinates. For Trip 1, 274 GPS observations were available (including the start and end points), with an average gap of $36$ minutes between two consecutive GPS observations. 130 GPS observations were available for Trip 2, with an average gap of $80$ minutes between two consecutive GPS observations. It is noteworthy that these GPS observations are not regularly spaced in time and the sample quantiles of the time gaps between two consecutive observations are summarized in Table~\ref{tb:quantilGaps}. For the DR path, the DR tag was originally programmed to sample at 16Hz, but the DRA was only performed for the thinned data at 1Hz, which was believed to be sufficiently fine for the purpose of tracking the animals~\citep{Benoit-Bird2013}.
\begin{table}[h]
\caption{The sample quantiles in minutes of the time gaps between two consecutive GPS observations in our case study.}\label{tb:quantilGaps}
\centering
\begin{tabular}{rrrrrrrr}
\toprule
 & Min & 10\% & 25\% & 50\% & 75\% & 90\% & Max \\ 
  \hline
Trip 1 & 14.75 & 15.00 & 15.45 & 18.40 & 31.68 & 82.79 & 953.65 \\ 
Trip 2 & 14.75 & 15.00 & 15.05 & 30.00 & 113.05 & 130.77 & 698.47 \\ 
\bottomrule
\end{tabular}
\end{table}

\subsection{BM analysis with a constant bias term}
As one illustrative example of our approach, we let $\bs{\beta} = \beta_0 = h(t)$, for the parametric bias part in~\eqref{eq:BM-DRModel}. The GPS observations of longitude and latitude are projected onto a plane in a point--wise fashion as in~\citet{Wilson2007}, such that the distance between any two consecutive GPS observations on this plane is their great circle distance and the angle between the line connecting the two points and the $y$--axis (latitude direction) equals the initial Bearing\footnote{Initial Bearing, also known as the forward azimuth, denotes the angle between the great circle connecting the two points on the earth and the north direction at the former point in time.} between them. The projected $x$--direction and $y$--direction in kilometres are called ``Easting" and ``Northing" to distinguish them from the longitude and latitude in degrees. As the two dimensions are analyzed separately, we have four data sets in total, which are referred to as ``Trip 1 Northing", ``Trip 1 Easting", ``Trip 2 Northing", and ``Trip 2 Easting". 

The GPS measurement variance in all of the analysis that follows was fixed at $\sigma^2_G= 0.0625$, which was chosen based on~\citet{Bryant2007} and the average observed number of satellites in those two trips. For the numerical integration part, $\delta_z=1$ and $\delta_\pi=3$ were set according to suggestions in~\citet{Rue2009}, which often led to around $33$ grid points in our data sets. All the computations required to obtain the posterior mean and variance for one data set can be done under 5 minutes of wall clock time on a regular laptop. The empirical Bayesian estimates of $\bs{\phi}$ are summarized in Table~\ref{tb:BM-EBayPhi-avgPostSE}.

\begin{figure}[h]
\includegraphics[width=\textwidth]{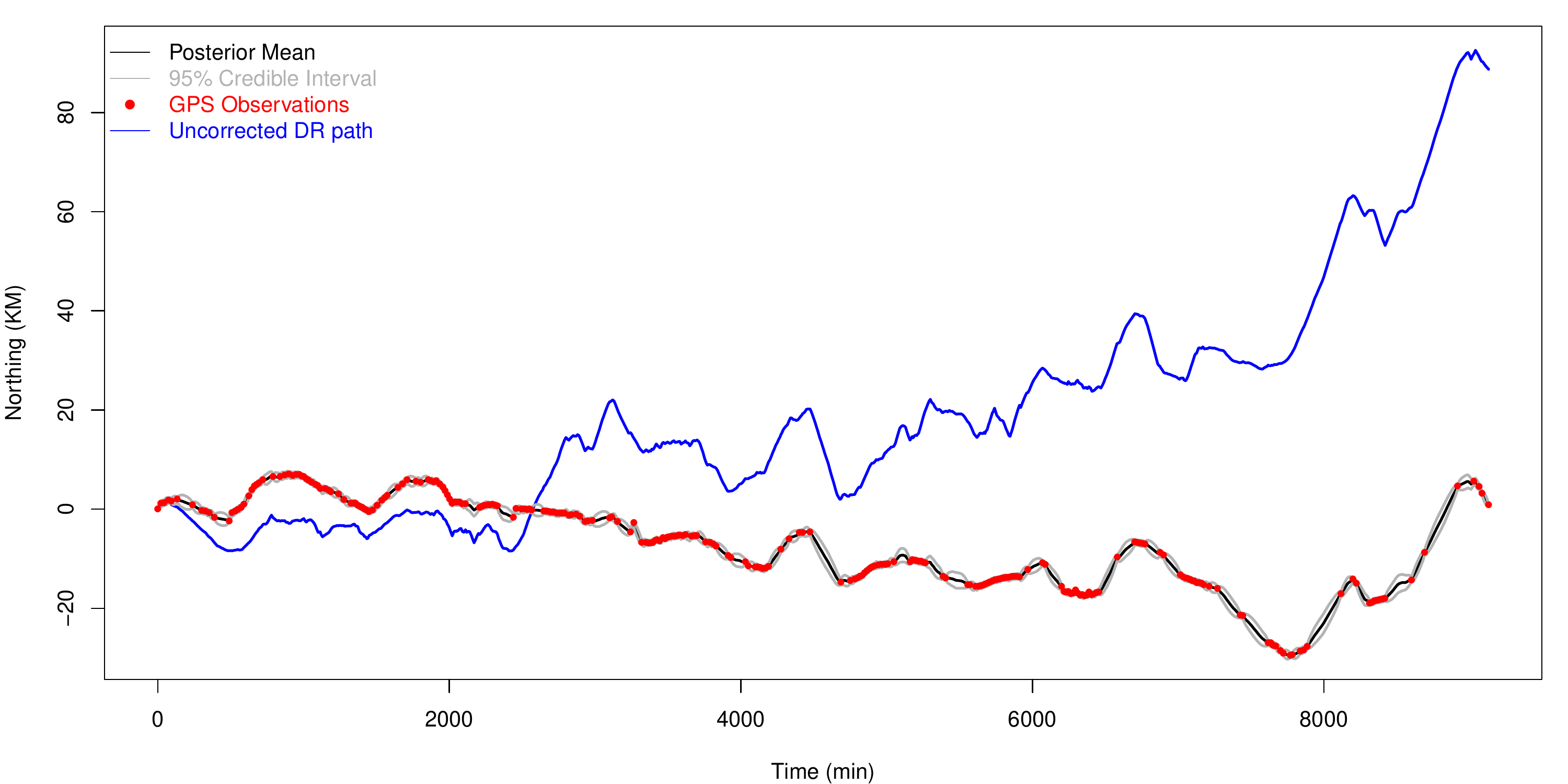}
\caption{The GPS observations and DRA outputs of the Trip 1 Northing data set together with the BM results for it. The red points are the GPS observations. The blue curve is the uncorrected DR path. The black curve is the posterior mean of $\bs{\eta}$ from our BM model. The grey curves connect the 95\% point--wise credible intervals at all the time points.}
 \label{fg:FG-SecBM-T1LatAll}
\end{figure}

In Figure~\ref{fg:FG-SecBM-T1LatAll}, we plot the GPS observations and DR path from Trip 1 latitude (Northing) together with our BM posterior mean and 95\% credible interval. The bias of the DR path dramatically increases with time and it is around 100KM at the end of this trip. Nonetheless, the DR path incorporates some fluctuations that match those seen in the fluctuations of the GPS observations. Our BM approach successfully moves the DR path results to the correct position taken to be the one indicated by the GPS points, while keeping the fluctuations of the DR path to reflect the animal's fine scale movements. 

\begin{figure}[h]
\includegraphics[width=\textwidth]{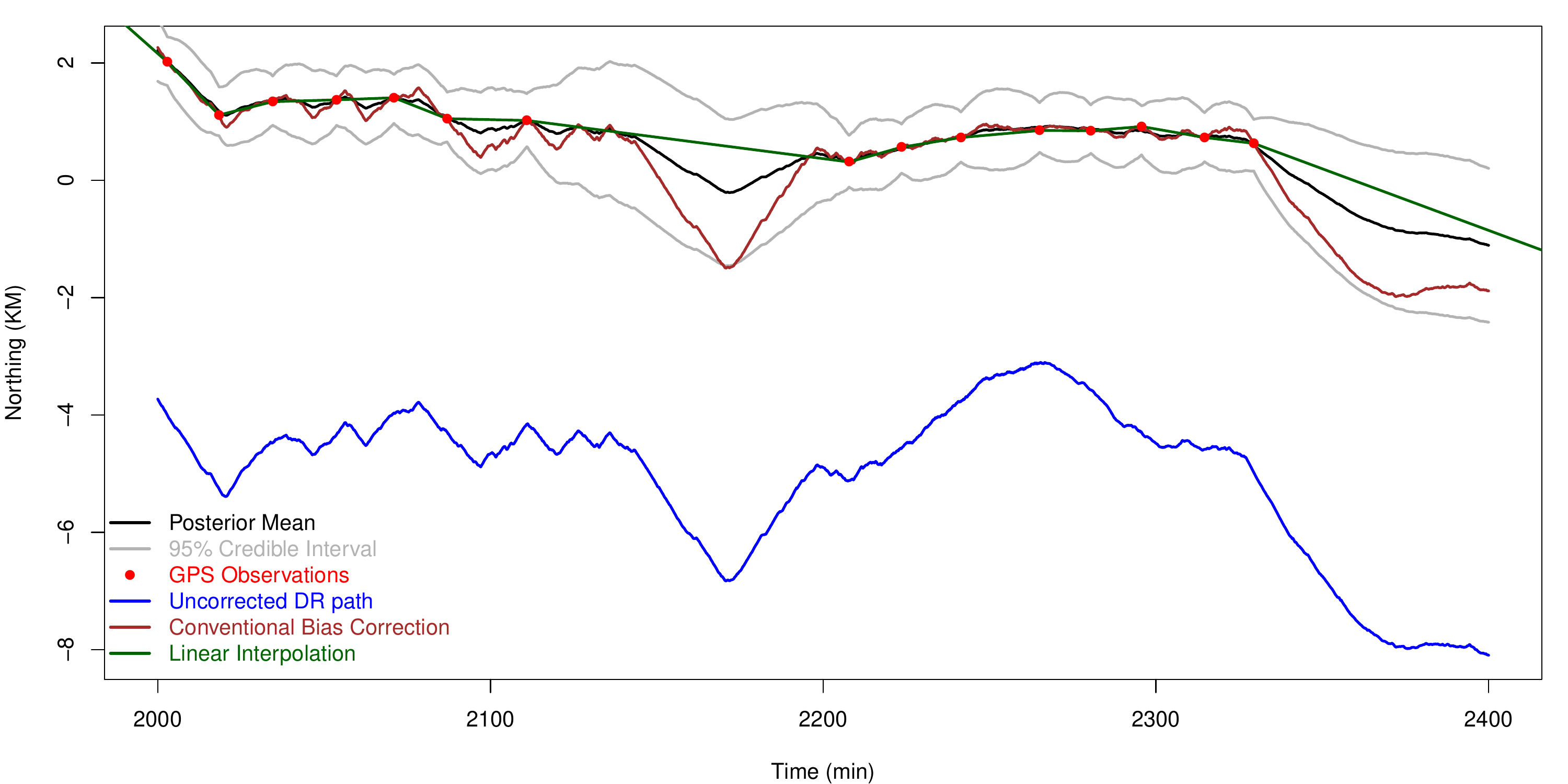} 
\caption{Zoom into the 2000-2400 mins part of Figure~\ref{fg:FG-SecBM-T1LatAll} (Trip 1 Northing data and BM results). The blue curve is the uncorrected DR path. The black curve is the posterior mean of $\bs{\eta}$ from our BM model. The grey curves connect the 95\% point-wise credible intervals at all the time points. The conventional bias correction in Equation~\eqref{eq:adhoc-DRABiasCorrection} is also included as the brown curve.} \label{fg:FG-SecBM-T1LatZoom}
\end{figure}

The scale of Figure~\ref{fg:FG-SecBM-T1LatAll} is a magnified plot that more fully displays how our approach works in the fine scale. Therefore, Figure~\ref{fg:FG-SecBM-T1LatZoom} zooms into the 2000--2400 minute portion of Figure~\ref{fg:FG-SecBM-T1LatAll}. The conventional bias correction from~\citet{Wilson2007} and linear interpolation between GPS observations are also included. The posterior mean from BM appears to be a shrunken version of the conventional bias correction, where the bumps in the conventional method are damped to the straight line connecting the two GPS points. This verifies our findings in Subsection~\ref{ss:theoryComparisonBMtoAdhoc}. 

Moreover, the CI for $\eta$ in Figure~\ref{fg:FG-SecBM-T1LatZoom} clearly displays a ``bridge" structure. Namely, the CI is narrow when $\eta(t)$ is close the GPS observations and becomes wider in between two GPS points. This is plausible as we have direct observations of the path at the GPS points but less accurate information from the DRA for points in between GPS points. Plots obtained by analyzing the other three data sets are quite similar to those seen here in substance and hence are omitted for brevity. 

In Table~\ref{tb:BM-EBayPhi-avgPostSE}, we also summarize the averaged posterior standard errors (APSEs) of the $\eta(t), t=1, 2, \ldots, T$ found in the four data sets. For the first trip, the APSE is around 500 meters in both dimensions, while it is around 1.3KM for the second trip. The differences in the averaged SEs are mainly decided by the gap between GPS observations. The longer the gap, the less accurate the corrected path is.
\begin{table}[h]
\caption{Empirical Bayesian estimates of $\bs{\phi}$ ($\hat{\sigma}^2_H, \hat{\sigma}^2_D$) and the averaged posterior standard error (APSE) in kilometre (KM) of $\eta(t)$ in the BM analyses of the four data sets from our case study.} \label{tb:BM-EBayPhi-avgPostSE}
\centering
\begin{tabular}{lccc}
\toprule
Data & $\hat{\sigma}^2_H$  & $\hat{\sigma}^2_D$ & APSE (KM) \\
\hline
Trip 1 Easting & 0.0801 & 0.0353 & 0.5243 \\ 
Trip 1 Northing& 0.0267 & 0.0428 & 0.4438 \\ 
Trip 2 Easting & 0.1031 & 0.0767 & 1.2880  \\
Trip 2 Northing& 0.1029 & 0.1233 & 1.4510\\
\bottomrule
\end{tabular}
\end{table}

\subsection{Cross validation comparisons} \label{ss:BM-RealDataAna-XV}
We also did a leave-one-out cross validation (LOOCV) analysis to further assess our BM approach. Here one GPS observation was deleted at a time when the model was trained. The difference between true observation and the prediction from our model was calculated and summarized as the cross validation root mean squared error (CV-RMSE). We also checked whether this observation is covered by the 95\% posterior CI of $\eta(t)$, which is summarized as the coverage percentage. The LOOCV was also performed for the conventional method and linear interpolation (connecting two consecutive GPS by a straight line), where only the CV-RMSE was obtained. The results are presented in Table~\ref{tb:BM-LOOCV-BMvsAD}. 

\begin{table}[h]
\caption{Results from the leave-one-out cross validation studies of the four data sets in our case study. The first column is the actual coverage percentage for the BM posterior 95\% CI. The cross validation root mean squared errors (CV-RMSE) in kilometres (KM) from our BM approach, conventional method from~\citet{Wilson2007} (Equation~\eqref{eq:adhoc-DRABiasCorrection}), and linear interpolation are in the last three columns.} \label{tb:BM-LOOCV-BMvsAD}
\vskip .1in
\centering
\begin{tabular}{l|c|c|c|c}
\toprule
				& \multicolumn{2}{c|}{BM Approach} & Conventional Method & Linear Interpolation \\
	Data Set	& Coverage & CV-RMSE & CV-RMSE & CV-RMSE \\	
\hline
Trip 1 Easting  & 99.3\%	& 0.374	& 0.485 & 0.433\\
Trip 1 Northing & 98.2\%	& 0.386	& 0.496 & 0.509\\
Trip 2 Easting  & 99.2\%	& 1.027 & 1.022 & 1.467\\
Trip 2 Northing & 100\%		& 0.849 & 1.148 & 1.061\\
\bottomrule
\end{tabular}
\end{table}

From Table~\ref{tb:BM-LOOCV-BMvsAD}, we see that the actual coverage of our 95\% CIs is higher than the nominal level, which are further discussed in~\ref{ss:BM-ConservativeCI}. On the other hand, it is easy to see that our BM approach has a smaller CV-RMSE than the linear interpolation method in all the four data sets while the conventional method has a larger CV-RMSE than the linear interpolation in two out of the four data sets, which indicates that the conventional method fails to use the information from the DR path appropriately. In this comparison of our BM to the conventional method, the CV-RMSE of our BM approach is smaller than those from the conventional method by around $1/3$ in three of the four data sets. For Trip 2 Easting, the CV-RMSE of our method is slightly larger than that of the conventional method. This might be caused by the fact that $h(t) = \beta_0$ failed to model the bias of DRA well and inspired us to consider different parametric models for $h(t)$ of~\eqref{eq:BM-DRModel}.

\subsection{Comparison of different models of $h(t)$ for the Trip 2 longitude data set} \label{ss:BM-T2Lon-ModelSelection}
Here polynomials $h(t) = \sum_{i=1}^Q \beta_i t^{i-1} $ up to order 6, ($Q=0, 1, 2, \ldots, 6$) were considered for Trip 2 Easting. The empirical Bayesian parameter estimates $\hat{\sigma}^2_H$ and $\hat{\sigma}^2_D$ under these models are in Table~\ref{tb:BM-ModelComp-T2Lon}. It is interesting to find that estimates of $\sigma^2_H$ almost stay the same when $Q$ increases. This is plausible as it is the parameter for the animal's Brownian Bridge movement, which is mainly decided by the GPS observations. On the other hand, the estimates of $\sigma^2_D$ decrease with increases in $Q$, as less randomness remains to be explained by the Brownian motion process when more flexible parametric models are characterized by $h(t)$. The smaller $\sigma^2_D$ also yields a larger $\rho = \frac{\sigma^2_H}{\sigma^2_H + \sigma^2_D}$, which indicates that more confidence can be placed on the DR path $X(t)$ and we thus get a more accurate $\eta(t)$. This explains why the posterior averaged SE decrease with increases in $Q$. 

We also compared the above models by means of the LOOCV and summarize the results in Table~\ref{tb:BM-ModelComp-T2Lon}. Clearly, the prediction accuracy of our model is improved by certain parametric models for $h(t)$. Models with $Q>1$ have smaller CV-RMSE than the conventional method ($1.022$ as from Table~\ref{tb:BM-LOOCV-BMvsAD}). When $Q=3$, the CV-RMSE is minimized. The coverage of the posterior CI varies slightly among those models and they are higher than the nominal coverage level.

\begin{table}[h]
\caption{Comparisons of different models for $h(t)$ in~\eqref{eq:BM-DRModel} for Trip 2 Easting (longitude) data. The first three columns of this table summarize the empirical Bayesian estimates of $\sigma^2_H, \sigma^2_D$ and the averaged posterior SE (APSE) in kilometre (KM). The last two columns are the cross-validation root mean squared error (CV-RMSE) and actual coverage percentage of the BM posterior 95\% CI in the leave-one-out cross validation.} \label{tb:BM-ModelComp-T2Lon}
\centering
\begin{tabular}{rrrrrr}
  \toprule
$Q$ & $\hat{\sigma}^2_H$  & $\hat{\sigma}^2_D$ & APSE (KM) & CV-RMSE & Coverage \\ 
  \hline
  0 & 0.1032 & 0.0827 & 1.314 & 1.035 & 0.992 \\ 
  1 & 0.1032 & 0.0767 & 1.288 & 1.027 & 0.992 \\ 
  2 & 0.1031 & 0.0693 & 1.252 & 1.018 & 0.984 \\ 
  3 & 0.1029 & 0.0458 & 1.101 & 0.969 & 0.984 \\ 
  4 & 0.1029 & 0.0462 & 1.104 & 0.969 & 0.984 \\ 
  5 & 0.1029 & 0.0447 & 1.093 & 0.981 & 0.992 \\ 
  6 & 0.1028 & 0.0379 & 1.035 & 1.005 & 0.977 \\ 
\bottomrule
\end{tabular}
\end{table}

\subsection{Are the CI's conservative? Leave-5-out Cross Validations} \label{ss:BM-ConservativeCI}
In the above LOOCV, the actual coverage percentage of the CI is higher than the nominal level, which brings the concern that the CI is conservative. However, LOOCV just evaluated the performance of our method in a short period without GPS observations,(e.g. mostly around 30 mins) while the time gaps between GPS observation in our case study can be much long than that, as shown in Table~\ref{tb:quantilGaps}. In order to assess the performance of our method in a longer period without GPS observations, we performed leave-5-out cross validations (L5OCV) of the same data sets. The gaps created by L5OCV were around 90min (Trip 1) and 150min (Trip 2), which are larger than the original 90\% sample quantiles of the gaps in Table~\ref{tb:quantilGaps}. Thus L5OCV offered a more realistic evaluation of our method. Similarly as in LOOCV, five observations were left out at a time when the model was trained the difference between the observations and our model predictions were calculated. The results are summarized in Table~\ref{tb:L5OCV}. From this table, we can easily verify that the coverage percentage are around the nominal level and thus the CI's from our BM are not conservative. The CV-RMSE's from Table~\ref{tb:L5OCV} once again demonstrate the superiority of our approach over the conventional method and linear interpolation.
\begin{table}[h!]
\caption{Results from the leave-five-out cross validation studies of the four data sets. The organization of this table is the same as Table~\ref{tb:BM-LOOCV-BMvsAD}.} \label{tb:L5OCV}
\vskip .1in
\centering
\begin{tabular}{l|c|c|c|c}
\toprule
	& \multicolumn{2}{c|}{BM Approach} & Conventional Method & Linear Interpolation\\
	Data Set	& Coverage & CV-RMSE & CV-RMSE & CV-RMSE\\	
\hline
Trip 1 Easting  & 97.8\%	& 0.726	& 1.039 & 1.131\\
Trip 1 Northing & 95.9\%	& 0.803	& 1.248 & 1.156\\
Trip 2 Easting  & 93.7\%	& 2.516 & 2.674 & 3.839\\
Trip 2 Northing & 92.9\%	& 3.061 & 3.326 & 4.436\\
\bottomrule
\end{tabular}
\end{table}

\section{Conclusions} \label{ss:BM-DiscussionFutureWork}
In this paper, we have developed a Brownian Bridge based Bayesian Melding approach to combine the sparse but accurate GPS observations with the high-resolution but biased DR path for the tracking of marine mammals. The posterior mean from our BM approach offers an accurate and high-resolution path of the tracked animals and the posterior credible intervals provides a reasonable statement of the uncertainty in our inferences. Moreover, our approach exploits the conditional independence property of the Brownian Bridge and Brownian Motion to dramatically reduce the heavy computational burden involved in dealing with large data sets. This work enables us to obtain a high--resolution in--situ record of the hydrographic data collected by the marine mammals, which may help to broaden our knowledge about parts of the ocean that are originally hard to observe and better address the recent changes in the global climate. Besides the contribution to environmental studies, our BM also can serve as a foundation for many biological and ecological questions such as the animal's habitat preference and resource selection~\citep{Hooten2013a}. Many aspects of the BM approach can be also further improved, such as analyzing the two dimensions of the path simultaneously and developing simultaneous credible intervals.

\vfill
\newpage
\begin{appendices}
\section{Details of the Dead-Reckoning Algorithm} \label{ap:DRA}
There are different versions of the DRA depending on the device and the subjects been tracked, but the main steps~\citep{Elkaim2006, Wilson2007}, can be summarized as follows:
\begin{enumerate}
\item Obtain the earth's gravity vector $\mb{g}$ and magnetic field vector $\mb{m}$ at the locations being studied.
\item Correct the systematic bias in the accelerometer and magnetometer~\citep{grewal2007global}.
\item Smooth the accelerometer and magnetometer readings by a running mean or a low-pass filter.
\item Find the animal's orientation (often taken to be the tag's orientation, namely the three-dimensional direction of velocity) by solving Wahba's problem~\citep{wahba1965least}, which is to solve for the rotation matrix $\mb{O}$ that minimizes the distance between the rotated gravity and magnetic force ($\mb{Og}$ and $\mb{Om}$) and the smoothed accelerometer reading $\mb{\tilde{a}}$ and magnetometer readings $\mb{\tilde{m}}$, namely,
\begin{align*}
\min_{\mb{O}} \left\{ || \mb{\tilde{a}} - \mb{Og} ||^2 + ||\mb{\tilde{m}} - \mb{Om}||^2 \right\}.
\end{align*}
\item Obtain the animal's speed (norm of the velocity vector) by one of the following approaches:
	\begin{enumerate}
	\item Assume it is a known constant. 
	\item Assume it is measured by a speed meter (a wheel or paddle), which could only measure the speed of the animal with respect to the water, but not to the earth. 
	\item From the data recorded by the TDR, calculate the velocity in the depth direction $v_z$. Given the animal's orientation $\mb{o}=\{o_x, o_y, o_z\}$, the three dimensional velocity is then $\mb{v} = \mb{o}v_z/o_z$.
	\end{enumerate}
\item Starting from a known point, integrate the velocity to obtain the animal's trajectory. 
\end{enumerate}
Although the DRA has been used for years and the data collecting devices are gradually being improved by electrical engineers, there is still huge bias in the DRA results. As shown later in this section, the bias of the DRA in our case study can be as large as 100km at the end of a seven-day trip. According to our analysis and some simulation studies, the bias of the DRA mainly owes to the following things:
\begin{enumerate}
\item Orientation bias: Although the gravity $\mb{g}$ can be precisely obtained, we can only obtain $\mb{m}$ on a relatively sparse grid on the earth at a certain time point~\citep{Wilson2007} rather than at detailed location and time points during the animal's trip. The imprecise magnetic vector introduces bias into the solution to Wahba's problem.

Moreover, an unbiased solution to the Wahba's problem (Step 4 above) requires the smoothed accelerometer reading $\tilde{\mb{a}}$ to be the projection of the gravity only. This is unlikely to be done by the smoother because the accelerometer reading is the projection of the sum of all the external forces, including the gravity, acceleration from the animal and forces from wave, etc. Similarly, the bias correction methods may fail to remove the systematic bias in the accelerometer and magnetometer readings~\citep{grewal2007global}.
\item Speed bias: The first two approaches above clearly involve some unrealistic assumptions. As for the third approach, it fails to work when the animal is floating on the surface ($o_z$ is zero). 
\item Discretization bias: Even if both biases above could be removed, we must still discretely approximate an integral in the continuous time domain. Although this problem can be alleviated by increasing the sampling frequency, that could reduce the tag's battery life and limit the scale of the study.
\end{enumerate}

\section{Details of the Bayesian Melding Model}
This section include the detailed derivations used in the inference of our BM model.
\subsection{Explicit expression of $\langle \bs{\beta}, \bs{\eta}|\mb{X}, \mb{Y}\rangle$} \label{ap:betaEta_Phi}
Following the model and notations in~\eqref{eq:BM-etaModel} to~\eqref{eq:BM-TargetPost}, it is easy to find
\begin{align}
\langle \bs{\beta}, \bs{\eta}|\mb{X}, \mb{Y}\rangle \propto & \langle \bs{\beta}, \bs{\eta}, \mb{X}, \mb{Y}\rangle = \langle \bs{\eta}\rangle \langle \mb{Y}|\bs{\eta}\rangle \langle \mb{X}|\bs{\beta}, \bs{\eta} \rangle \notag \\
 \propto& |\mb{R}|^{-1/2} \exp\left\{- \frac{1}{2} (\bs{\eta} - \mb{f})^T \mb{R}^{-1} (\bs{\eta} - \mb{f}) \right\} \times
|\mb{D}|^{-1/2} \exp\left\{ - \frac{1}{2} (\mb{Y} - \mb{G}\bs{\eta} )^T \mb{D}^{-1}(\mb{Y}-\mb{G}\bs{\eta} ) \right\} \notag \\
& \times |\mb{C}|^{-1/2} \exp\left\{ - \frac{1}{2} (\mb{X} - \mb{Z}\bs{\beta} - \bs{\eta})^T \mb{C}^{-1}(\mb{X} - \mb{Z}\bs{\beta} - \bs{\eta}) \right\} \label{eq:BM-cPhi-Step1},
\end{align}
where the following notations are used in addition to those from~\eqref{eq:BM-etaModel} to~\eqref{eq:BM-TargetPost}:
\begin{itemize}
\item $\mb{f} =\{f(2), f(3), \ldots, f(T-1) \}^T$, a $T-2$ vector as the prior mean of $\bs{\eta}$;
\item $\mb{D} = \sigma^2_G \mb{I}_{K-2}$ is the covariance matrix of $\mb{Y}$ conditioning on $\bs{\eta}$, where $\mb{I}_{m}$ stands for the identity matrix of dimension $m$; 
\item $\mb{Z}$ the design matrix for $\mb{h}$ as in~\eqref{eq:BM-DRModel}
\item $\mb{G}$ is a $(K-2)\times (T-2)$ matrix, with
	\begin{align*}
	G_{k, j} = \left\{ \begin{aligned}
	1, & & j=t_k -1 \\
	0, & & \text{Otherwise}
	\end{aligned}\right.
	\end{align*}
	Notice here the first time point is removed;
\item $\mb{R} = \mb{R}(2:(T-1), 2:(T-1))$ is the $(T-2) \times (T-2)$ covariance matrix for the Brownian Bridge;
\item $\mb{C} = \mb{C}(2:T, 2:T)$ is the $(T-1) \times (T-1)$ covariance matrix for the Brownian Motion error term $\bs{\xi}$;
\end{itemize} 
In order to further simplify~\eqref{eq:BM-cPhi-Step1}, the following notations are introduced:
\begin{itemize}
\item $\bs{\zeta} = \{\bs{\beta}^T, \bs{\eta}^T\}^T$, which is the joint vector of $\bs{\beta}$ and $\bs{\eta}$ of length $Q + T-2$;
\item $\mb{U}$ is a $T-1 \times (Q+T-2)$ matrix with $\mb{U}(1:(T-1), 1:Q) = \mb{Z}$, $\mb{U}(1:(T-2), (Q+1):(Q+T-2)) = I_{T-2}$ and  $\mb{U}(T-1, (Q+1):(Q+T-2)) = 0$, which maps $\mb{U}\bs{\zeta} = \mb{h} + \bs{\beta}$;
\item $\mb{V}$ is a $(T-2)\times (Q+T-2)$ matrix with $\mb{V}(1:(T-2), 1:Q) = \mb{0}$ and $\mb{V}(1:(T-2), (Q+1):(Q+T-2)) = \mb{I}_{T-2}$, such that $\mb{V}\bs{\zeta} = \bs{\eta}$;
\item $\mb{W}$ is a $(K-2)\times (Q+T-2)$ matrix  with $\mb{W}(1:(K-2), 1:Q) = 0$ and $\mb{W}(1:(K-2), (Q+1):(Q+T-2)) = \mb{G}$, such that $\mb{W}\bs{\zeta} = \mb{G}\bs{\eta}$.
\end{itemize}

Some algebra could simplify~\eqref{eq:BM-cPhi-Step1} into 
\begin{align}
\langle \bs{\zeta}|\mb{X}, \mb{Y}\rangle  \propto &|\mb{R}|^{-1/2}|\mb{D}|^{-1/2}|\mb{C}|^{-1/2} \times \notag \\
&\exp\left\{ -\frac{1}{2} \left[ (\bs{\zeta}-\mb{M}^{-1}_1 \mb{M}_2)^T \mb{M}_1 (\bs{\zeta}-\mb{M}^{-1}_1 \mb{M}_2)  + M_3 -\mb{M}^T_2 \mb{M}^{-1}_1 \mb{M}_2 \right] \right\}, \notag \\
\propto& \exp\left\{ -\frac{1}{2} \left[ (\bs{\zeta}-\mb{M}^{-1}_1 \mb{M}_2)^T \mb{M}_1 (\bs{\zeta}-\mb{M}^{-1}_1 \mb{M}_2) \right] \right\}
 \label{eq:BM-cPhi-Step2}
\end{align}
where
\begin{align*}
\mb{M}_1 =& \mb{V}^T \mb{R}^{-1}\mb{V} + \mb{W}^T\mb{D}^{-1}\mb{V}_G  + \mb{U}^T\mb{C}^{-1}\mb{U} \\
\mb{M}_2 =& \mb{V}^T \mb{R}^{-1}\mb{f}  + \mb{W}^T\mb{D}^{-1}\mb{Y} + \mb{U}^T\mb{C}^{-1}\mb{X} \\
M_3 =& \mb{f}^T\mb{R}^{-1}\mb{f} + \mb{Y}^T \mb{D}^{-1}\mb{Y} + \mb{X}^T\mb{C}^{-1}\mb{X}.
\end{align*}
It is shown here that the posterior of $\bs{\beta}, \bs{\eta}$ conditioning on $\bs{\phi}$ is a multivariate Gaussian density. The main computation complexity of evaluating this density comes from calculation of $\mb{M}_1^{-1}\mb{M}_2$.

\subsection{Derivation of~\eqref{eq:BM-cPhi-b2} and~\eqref{eq:BM-cPhi-b3}} \label{ap:condIndep}
It is well known that the Brownian Bridge and Brownian Motion process are Markovian~\citep{stirzaker2001probability}, such that: 
\begin{align*}
[\eta(t)|\eta(t-1), \eta(t-2)] =& [\eta(t)|\eta(t-1)] \\
[\xi(t)|\xi(t-1), \xi(t-2)] =& [\xi(t)|\xi(t-1)],
\end{align*}
where $\eta(t)$ is a Brownian Motion process as in our model~\eqref{eq:BM-etaModel} and $\xi(t)$ is a Brownian Motion process as in~\eqref{eq:BM-DRModel}. This Markovian property directly suggests the conditional independence property of these two process, such that:
\begin{align*}
[\eta(t-1, t+1)|\eta(t)] =& [\eta(t-1)|\eta(t)][\eta(t+1)|\eta(t)]\\
[\xi(t-1, t+1)|\xi(t)] =& [\xi(t-1)|\xi(t)][\xi(t+1)|\xi(t)].
\end{align*}
These properties help us to derive~\eqref{eq:BM-cPhi-b2} and~\eqref{eq:BM-cPhi-b3}. As an illustration, we assume $T=5$ (recall that $\eta(1)$ and $\eta(5)$ are fixed. So only $\bs{\eta}(2, 3, 4)$ are random), and one GPS observation is available at $t=3$. The first part of~\eqref{eq:BM-cPhi-b1}, $\langle \bs{\eta}(1:T\setminus t_{1:K})|\bs{\beta}, \bs{\eta}_G, \mb{X}, \mb{Y} \rangle$ under this situation is 
\begin{align}
\langle\eta(2), \eta(4)|\eta(3), \mb{X}(2:4), Y(3), \bs{\beta} \rangle = \langle\eta(2), \eta(4)|\eta(3), \mb{X}(2:4), \bs{\beta}\rangle, \label{eq:BM-CondInd-Ex1}
\end{align}
as $Y(3)$ only depends $\eta(3)$ and independent of all the other random variables. For brevity, we will hide $\bs{\beta}$ term in all the following derivation and let $X_i = X(i), \eta_i = \eta(i), \xi_i= \xi(i)$. Using the conditional independence property of $\eta$ and $\xi$ together with certain variable transformation, we can simplify~\eqref{eq:BM-CondInd-Ex1} into 
\begin{align}
\langle\eta_2, \eta_4|\eta_3, X_{2:4}\rangle &= \frac{\langle\eta_2, \eta_3, \eta_4, X_2, X_3, X_4\rangle}{\langle \eta_3, X_2, X_3, X_4\rangle} = \frac{\langle\eta_2, \eta_3, \eta_4, \xi_2, \xi_3, \xi_4\rangle}{\langle \eta_3, X_2, X_3, X_4\rangle} \notag \\
&= \frac{\langle\eta_2, \eta_3, \eta_4\rangle \langle \xi_2, \xi_3, \xi_4\rangle}{\langle \eta_3, X_2, X_3, X_4\rangle} =  \frac{\langle\eta_2|\eta_3\rangle \langle \eta_4|\eta_3\rangle \langle \eta_3 \rangle \langle\xi_2|\xi_3\rangle \langle \xi_4|\xi_3\rangle \langle \xi_3 \rangle}{\langle \eta_3, X_2, X_3, X_4\rangle } \notag \\
&= \frac{\langle\eta_2|\eta_3\rangle \langle\xi_2|\xi_3\rangle \langle \eta_4|\eta_3\rangle \langle \xi_4|\xi_3\rangle  \langle \eta_3 \rangle \langle \xi_3 \rangle }{\langle \eta_3, X_2, X_3, X_4\rangle} \notag \\
&= \frac{\langle\eta_2, X_2|\eta_3, X_3\rangle\langle \eta_4, X_4|\eta_3, X_3\rangle \langle \eta_3, X_3\rangle }{\langle \eta_3, X_2, X_3, X_4\rangle} \notag \\
&= \frac{\langle\eta_2, X_2|\eta_3, X_3\rangle\langle \eta_4, X_4|\eta_3, X_3\rangle}{\langle X_2, X_4|\eta_3, X_3 \rangle} = \ldots = \frac{\langle\eta_2, X_2|\eta_3, X_3\rangle\langle \eta_4, X_4|\eta_3, X_3\rangle}{\langle X_2|\eta_3, X_3 \rangle\langle X_4|\eta_3, X_3 \rangle}  \notag \\
&= \langle\eta_2|\eta_3, X_2, X_3\rangle \langle\eta_4|\eta_3, X_4, X_3\rangle. \notag
\end{align}
The above is an illustration on how we prove~\eqref{eq:BM-cPhi-b2} and~\eqref{eq:BM-cPhi-b3} and we omit the lengthy proof of the general case. On the other hand, these expressions can be easily proved via the conditional independence property of a graphical model as in~\citet{lauritzen1996graphical}.

\subsection{Explicit expression of~\eqref{eq:BM-cPhi-b3}} \label{ap:BM-perCond}
In the above subsection, we have shown that:
\begin{align*}
\langle \bs{\eta}(t_k +1 : t_{k+1}-1)|& \eta(t_k), \eta(t_{k+1}), \bs{\beta}, \mb{X}, \mb{Y} \rangle =  \\
&\langle \bs{\eta}(t_k +1 : t_{k+1}-1)| \eta(t_k), \eta(t_{k+1}), \bs{\beta}, \mb{X}(t_k : t_{k+1})\rangle.
\end{align*}
Here we derive the explicit expression of the right-hand side above. First, it is easy to use the conditional independence property to find:
\begin{align}
\langle \bs{\eta}(t_k +1 : t_{k+1}-1)|& \eta(t_k), \eta(t_{k+1}), \bs{\beta}, \mb{X}(t_k : t_{k+1})\rangle
 \notag \\ =& \frac{\langle \bs{\eta}(t_k +1 : t_{k+1}-1)| \eta(t_k), \eta(t_{k+1}), \bs{\beta}, X(t_k),  X(t_{k+1})\rangle }{ \langle \mb{X}(t_k +1 : t_{k+1}-1) |\eta(t_k), \eta(t_{k+1}), \bs{\beta}, X(t_k),  X(t_{k+1}) \rangle} \notag \\
=&  \frac{\langle \bs{\eta}(t_k +1 : t_{k+1}-1)| \eta(t_k), \eta(t_{k+1})\rangle }{ \langle \mb{X}(t_k +1 : t_{k+1}-1) |\eta(t_k), \eta(t_{k+1}), \bs{\beta}, X(t_k),  X(t_{k+1}) \rangle}. \notag
\end{align}
Define two new variables,
\begin{align*}
\bs{\eta}^c(t_k+1: t_{k+1}-1) =& \bs{\eta}(t_k +1 : t_{k+1}-1)| \eta(t_k), \eta(t_{k+1}) \\
\mb{X}^c(t_k+1: t_{k+1}-1) =& \mb{X}(t_k +1 : t_{k+1}-1) |\eta(t_k), \eta(t_{k+1}), \bs{\beta}, X(t_k),  X(t_{k+1}),
\end{align*}
such that
\begin{align}
\langle \bs{\eta}(t_k +1 : t_{k+1}-1)|& \eta(t_k), \eta(t_{k+1}), \bs{\beta}, \mb{X}(t_k : t_{k+1})\rangle \notag \\
 =& \langle \bs{\eta}^c(t_k+1: t_{k+1}-1)| \mb{X}^c(t_k+1: t_{k+1}-1) \rangle \label{eq:BM-perCond-EtaC_XC}
\end{align}
Accoridng to the basic properties of the Brownian Bridge, it is easy to verify that
\begin{align}
\bs{\eta}^c(t_k+1: t_{k+1}-1) \sim& \text{MVN}(\mb{f}_k, \sigma^2_H \mb{R}_k), \label{eq:BM-perCond-EtaC-Def}
\end{align}
where 
\begin{align*}
f_k(t) =& f(t) + a'_k(t)(\eta(t_k) - f(t_k)) + a_k(t)(\eta(t_{k+1}) - f(t_{k+1})) \\
 =& a'_k(t)\eta(t_k) +  a_k(t)\eta(t_{k+1})\\
a_k(t) =& \frac{t - t_k}{t_{k+1} - t_k}\\
a'_k(t) =& 1- a_k(t) = \frac{t_{k+1} - t}{t_{k+1} - t_k}\\
R_k(s, t) =&  \frac{(s - t_k)(t_{k+1}-t)}{t_{k+1} - t_k}, &  t_k <s \leq t<  t_{k+1},
\end{align*}
and $\mathrm{MVN}(\cdot, \cdot)$ denotes a multivariate normal distribution.

The definition of $\mb{X}^c$ and $\mb{X}$ in~\eqref{eq:BM-DRModel} implies that $\mb{X}^c$ is the sum of a deterministic term and two independent Gaussian processes:
\begin{align*}
\mb{X}^c(t_k+1: t_{k+1}-1) = \mb{h}(t_k+1: t_{k+1}-1) + \bs{\eta}^c(t_k+1: t_{k+1}-1) +  \bs{\xi}^c(t_k+1: t_{k+1}-1),
\end{align*} 
where $\mb{h}$ is defined in~\eqref{eq:BM-DRModel} and $\bs{\xi}^c$ is defined similarly as $\bs{\eta}^c$:
\begin{align*}
\bs{\xi}^c(t_k+1: t_{k+1}-1) = \bs{\xi}(t_k +1 : t_{k+1}-1)| \xi(t_k), \xi(t_{k+1}) \sim \text{MVN}(\mb{g}_k, \sigma^2_D \mb{R}_k)
\end{align*}
with $g_k(t) = a'_k(t)\xi(t_k) + a_k(t)\xi(t_{k+1})$.

In this way, the marginal distribution of $\mb{X}^c$ is
\begin{align}
\mb{X}^c(t_k+1: t_{k+1}-1) \sim \text{MVN}(\mb{u}_k, (\sigma^2_H + \sigma^2_D)\mb{R}_k), \label{eq:BM-perCond-XC-Def}
\end{align}
where 
\begin{align*}
u_k(t) &=h(t) + f_k(t) + g_k(t)\\
 &=h(t) +  a'_k(t) (\eta(t_k) + \xi(t_k)) + a_k(t)(\eta(t_{k+1})+\xi(t_{k+1})) \\
 &=h(t) + a'_k(t) (X(t_k) - h(t_k)) + a_k(t)(X(t_{k+1}) - h(t_{k+1})).
\end{align*}
Also, the covariance between $\bs{\eta}^c$ and $\mb{X}^c$ is 
\begin{align}
\Cov(\bs{\eta}^c, \mb{X}^c) = \Cov(\bs{\eta}^c, \bs{\eta}^c) = \sigma^2_H \mb{R}_k. \label{eq:BM-perCond-CovEtaCXC-Def}
\end{align}

With~\eqref{eq:BM-perCond-EtaC-Def},~\eqref{eq:BM-perCond-XC-Def}, and~\eqref{eq:BM-perCond-CovEtaCXC-Def}, we can easily find that there is no need to invert any matrices when computing~\eqref{eq:BM-perCond-EtaC_XC}. As when the conditional covariance is calculated, the inverse of $\mb{R}_k$ will be canceled by $\mb{R}_k$, such that,
\begin{align*}
\Cov(\bs{\eta}^c, \mb{X}^c) \left(\Cov(\mb{X}^c) \right)^{-1}= &\sigma^2_H \mb{R}_k \left( (\sigma^2_H + \sigma^2_D)\mb{R}_k) \right)^{-1} =  \frac{\sigma^2_H}{\sigma^2_H + \sigma^2_D} \mb{I}_{t_{k+1} - t_k-1} \\
\Cov(\bs{\eta}^c, \mb{X}^c) \left(\Cov(\mb{X}^c) \right)^{-1}\Cov( \mb{X}^c, \bs{\eta}^c) =& \sigma^2_H(1 -  \frac{\sigma^2_H}{\sigma^2_H + \sigma^2_D})\mb{R}_k = \frac{\sigma^2_H\sigma^2_D}{\sigma^2_H + \sigma^2_D}\mb{R}_k.
\end{align*} 
We only need to calculate $ \rho = \frac{\sigma^2_H}{\sigma^2_H + \sigma^2_D}$ for the conditional mean and covariance. In this way, the desired posterior in~\eqref{eq:BM-perCond-EtaC_XC} is 
\begin{align}
\langle \bs{\eta}(t_k +1 : t_{k+1}-1)|&\eta(t_k), \eta(t_{k+1}), \bs{\beta}, \mb{X}(t_k : t_{k+1})\rangle \notag \\ =&\langle \bs{\eta}^c(t_k+1: t_{k+1}-1)| \mb{X}^c(t_k+1: t_{k+1}-1) \rangle \notag \\
\sim& \text{MVN}\left(\mb{f}_k +\rho\left(  \mb{X}(t_k+1: t_{k+1}-1) - \mb{u}_k \right), \rho \sigma^2_D \mb{R}_k \right) \label{eq:BM-perCond-EtaNG_ZetaGPhiXY}
\end{align}

\subsection{Explicit expression of $[\bs{\phi}, \bs{\beta}, \bs{\eta}_G|\mb{X}_G, \mb{Y}]$} \label{ap:betaEtaGPhi_XGY}
Following notations in Appendix~\ref{ap:betaEta_Phi}
\begin{align}
[\bs{\phi}, \bs{\beta}, & \bs{\eta}_G|\mb{X}_G, \mb{Y}] \propto [\bs{\phi}][\bs{\eta}(t_{1:K})|\bs{\phi}] [\mb{Y}|\bs{\eta}(t_{1:K})] [\mb{X}(t_{1:K})|\bs{\beta}, \bs{\eta}(t_{1:K}), \bs{\phi}]\notag \\
 \propto&[\bs{\phi}] |\mb{R}_G|^{-1/2} \exp\left\{- \frac{1}{2} (\bs{\eta}_G - \mb{f}_G)^T \mb{R}_G^{-1} (\bs{\eta}_G - \mb{f}_G) \right\} \times
|\mb{D}|^{-1/2} \exp\left\{ - \frac{1}{2} (\mb{Y} - \bs{\eta}_G )^T \mb{D}^{-1}(\mb{Y}-\bs{\eta}_G) \right\} \notag \\
& \times |\mb{C}_G|^{-1/2} \exp\left\{ - \frac{1}{2} (\mb{X}_G - \mb{Z}_G\bs{\beta} - \bs{\eta}_G)^T \mb{C}_G^{-1}(\mb{X}_G - \mb{Z}_G\bs{\beta} - \bs{\eta}_G) \right\} \label{eq:BM-GBeta-step1},
\end{align}
where $\mb{X}_G= \mb{X}(t_{1:K})$ and $\mb{R}_G= \mb{R(t_{1:K}, t_{1:K})}$. Similar notations apply to $\bs{\eta}_G$, $\mb{f}_G, \mb{C}_G, \mb{Z}_G$. Also, we will introduce the following notations similar to those  in Appendix~\ref{ap:betaEta_Phi}, which is the sub-vector or sub-matrix of those in Appendix~\ref{ap:betaEta_Phi} with respect to the GPS observations:
\begin{itemize}
\item $\bs{\zeta}_G = \{\bs{\beta}^T, \bs{\eta}_G^T\}^T$, which is the joint vector of $\bs{\beta}$ and $\bs{\eta}_G$ of length $Q + K-2$;
\item $\mb{U}_G$ is a $K-1 \times (Q+K-2)$ matrix with $\mb{U}_G(1:(K-1), 1:Q) = \mb{Z}_G$, $\mb{U}_G(1:(K-2), (Q+1):(Q+T-2)) = I_{K-2}$ and  $\mb{U}_G(K-1, (Q+1):(Q+T-2)) = 0$, which maps $\mb{U}_G\bs{\zeta}_G = \mb{h}_G + \bs{\beta}_G$;
\item $\mb{V}_G$ is a $(K-2)\times (Q+K-2)$ matrix with $\mb{V}_G(1:(K-2), 1:Q) = \mb{0}$ and $\mb{V}_G(1:(K-2), (Q+1):(Q+K-2)) = \mb{I}_{K-2}$, such that $\mb{V}_G\bs{\zeta}_G = \bs{\eta}_G$;
\end{itemize}

Some algebra could simplify~\eqref{eq:BM-GBeta-step1} into 
\begin{align}
[\bs{\phi}, \bs{\beta}, & \bs{\eta}_G|\mb{X}_G, \mb{Y}] \propto \notag \\ &\frac{1}{\sigma^2_H} \frac{1}{\sigma^2_D} \times |\mb{R}_G|^{-1/2}|\mb{D}|^{-1/2}|\mb{C}_G|^{-1/2} \times \notag \\
&\exp\left\{ -\frac{1}{2} \left[ (\bs{\zeta}_G-\mb{M}^{-1}_{G1} \mb{M}_{G2})^T \mb{M}_{G1} (\bs{\zeta}_G-\mb{M}^{-1}_{G1} \mb{M}_{G2})  + M_{G3} -\mb{M}^T_{G2} \mb{M}^{-1}_{G1} \mb{M}_{G2} \right] \right\}, \label{eq:BM-GBeta-step2}
\end{align}
where
\begin{align*}
\mb{M}_{G1} =& \mb{V}_G^T \mb{R}_G^{-1}\mb{V}_G + \mb{V}_G^T\mb{D}^{-1}\mb{V}_G  + \mb{U}_G^T\mb{C}^{-1}\mb{U}_G \\
\mb{M}_{G2} =& \mb{V}_G^T \mb{R}_G^{-1}\mb{f}_G + \mb{V}_G^T\mb{D}^{-1}\mb{Y} + \mb{U}_G^T\mb{C}_G^{-1}\mb{X}_G \\
M_{G3} =& \mb{f}_G^T\mb{R}_G^{-1}\mb{f}_G + \mb{Y}^T \mb{D}^{-1}\mb{Y} + \mb{X}_G^T\mb{C}_G^{-1}\mb{X}_G.
\end{align*}

Following~\eqref{eq:BM-GBeta-step2}, it is easy to learn that
\begin{align}
[\bs{\zeta}_G|\mb{X}_G, \mb{Y}, \bs{\phi}] \sim MVN \left(\mb{M}^{-1}_{G1} \mb{M}_{G2}, \mb{M}^{-1}_{G1}\right). \label{eq:BM-ZetaG_PhiXY}
\end{align}
Integrate $\bs{\zeta}_G$ out, we have
\begin{align}
[\bs{\phi}|\mb{X}_G, \mb{Y}] \propto \frac{1}{\sigma^2_H} \frac{1}{\sigma^2_D} |\mb{R}_G|^{-1/2}|\mb{D}|^{-1/2}|\mb{C}_G|^{-1/2} |\mb{M}_{G1}|^{-1/2} \exp\left\{ -\frac{1}{2} \left[M_{G3} -\mb{M}^T_{G2} \mb{M}^{-1}_{G1} \mb{M}_{G2} \right] \right\}
 \end{align}

\subsection{Marginal distribution of $\bs{\eta}$ at the non-GPS points} \label{ap:EtaNonG_XY}
With~\eqref{eq:BM-ZetaG_PhiXY}, we can marginalize $\bs{\eta}_G, \bs{\beta}$ out  in~\eqref{eq:BM-perCond-EtaNG_ZetaGPhiXY} to obtain $\langle \bs{\eta}(t_k +1 : t_{k+1}-1)| \mb{X}, \mb{Y}\rangle$, which is later used in the numerical integration. Let $\bs{\zeta}_k =\{\bs{\beta}^T, \eta(t_k), \eta(t_{k+1}) \}^{T}$ and its mean and covariance matrix obtained in~\eqref{eq:BM-ZetaG_PhiXY} be denoted as $\bs{\tilde{\zeta}}_k, \bs{\tilde{\Sigma}}_k$ respectively.

The $\mb{u}_k$ of~\eqref{eq:BM-perCond-XC-Def} can be written as as a linear transformation of $\bs{\zeta}_k$, such that $\mb{u}_k = \mb{B}_k \bs{\zeta}_k$, where $\mb{B}_k = \left[\rho \left( \mb{Z}_k - \mb{A}_k  \mb{Z}^G_k \right), \mb{A}_k \right]$. $\mb{Z}_k$ is the rows of the design matrix, $\mb{Z}(t_k+1:t_{k+1}-1, 1:Q)$, which corresponds to this period of the non-GPS observation. $\mb{Z}^G_k = \mb{Z}(\mb{t}_{k, k+1}, 1:Q)$ corresponds to the rows of the design matrix of the two GPS observations. $\mb{A}_k$ is the matrix of the linear weights of $a'_k(t), a_k(t)$ as in Equation~\eqref{eq:BM-perCond-EtaC-Def}. Marginalize $\bs{\zeta}_k$ out in~\eqref{eq:BM-perCond-EtaNG_ZetaGPhiXY} results in 
\begin{align}
\langle \bs{\eta}(t_k +1 : t_{k+1}-1)| \mb{X}, \mb{Y}\rangle \sim \text{MVN}\left( \mb{B}_k\bs{\tilde{\zeta}}_k , \mb{B}_k\bs{\tilde{\Sigma}}_k\mb{B}^T_k \right) \label{eq:BM-ZetaNG_PhiXY}
\end{align}  

\subsection{Integration part of~\eqref{eq:BM-JointPostExpression} via a Gaussian mixture} \label{ap:IntegrationAsNormalMixture}
As introduced in Subsection~\ref{ss:IntegrationOfPhi}, we evaluate the integration part of~\eqref{eq:BM-JointPostExpression} via the numerical integration in~\eqref{eq:BM-etaPost-GridApprxoimation}. 
The distribution of $\bs{\zeta}$ is multivariate normal conditioning on $\bs{\phi}$ as  in~\eqref{eq:BM-ZetaG_PhiXY} and~\eqref{eq:BM-ZetaNG_PhiXY} and therefore the posterior density of $\bs{\zeta}$ can be approximated by a mixture of multivariate normal densities. Let $\tilde{\bs{\zeta}}^{(i)}$ and $\bs{\tilde{\Sigma}}^{(i)}$ be the posterior mean and covariance of $\bs{\zeta}$ conditioning on the $i$-th grid point of $\bs{\phi}(\mb{z}(j_1, j_2))$ (Here $j_1, j_2$ are collapsed into a single index set and $L= (J^{+}_1 + J^{-}_1 + 1) \times (J^{+}_2 + J^{-}_2 + 1)$). We simplify~\eqref{eq:BM-etaPost-GridApprxoimation} into 
\begin{align*}
[\bs{\zeta}|\mb{X}, \mb{Y}] \approx \sum_{i=1}^L w_i [\bs{\zeta}|\bs{\phi}^{(i)}, \mb{X}, \mb{Y}] 
=\sum_{i=1}^L w_i \Psi(\cdot; \bs{\tilde{\zeta}}^{(i)}, \bs{\tilde{\Sigma}}^{(i)}),
\end{align*} 
where $\bs{\Psi}$ stands for the probability density function of the multivariate normal distribution. For our application, we are only interested in finding out the posterior mean and variance of $\bs{\zeta}$. As in~\citep{fruhwirth2006finite}, the posterior mean equals 
\begin{align*}
\bs{\tilde{\zeta}} = \sum_{i=1}^L w_i  \bs{\tilde{\zeta}}^{(i)}, 
\end{align*}
and the posterior variance of the $k$-th element of $\bs{\zeta}$ is 
\begin{align*}
\tilde{\sigma}^2_k = \sum_{i=1}^L w_i \left[\tilde{\Sigma}^{(i)}(k, k) + \sum_{i=1}^L \left(\tilde{\zeta}^{(i)}(k) -\tilde{\zeta}(k) \right)^2\right].
\end{align*}
\end{appendices}
\bibliographystyle{apalike}
\bibliography{library,booksBib} 
\end{document}